\documentclass[runningheads]{llncs}
\usepackage[T1]{fontenc}
\usepackage[colorlinks,linkcolor=blue]{hyperref}
\usepackage{multirow}
\usepackage{graphicx}
\usepackage{amsmath}
\usepackage{bbm}
\usepackage{amssymb}
\usepackage{xspace}
\newcommand{\ie}{i.e.\@\xspace}

\newcommand{\eg}{e.g.\@\xspace}

\begin{document}
\title{Poisson Ordinal Network for Gleason Group Estimation Using Bi-Parametric MRI}
\titlerunning{Poisson Ordinal Network}

\author{
Yinsong~Xu\inst{1,2}  \and % index{Xu, Yinsong}
Yipei~Wang\inst{1}  \and % index{Wang, Yipei}
Ziyi~Shen\inst{1}  \and % index{Shen, Ziyi}
Iani~J.M.B.~Gayo\inst{1} \and % index{Gayo, Iani}
Natasha~Thorley\inst{3} \and % index{Thorley, Natasha}
Shonit~Punwani\inst{3} \and% index{Punwani, Shonit}
Aidong~Men\inst{2} \and% index{Men, Aidong}
Dean~Barratt\inst{1} \and% index{Barratt, Dean}
Qingchao~Chen\inst{4} \and% index{Chen, Qingchao}
Yipeng~Hu\inst{1}} % index{Hu, Yipeng}
\institute{Centre for Medical Image Computing, University College London, London, UK \and
School of Artificial Intelligence, Beijing University of Posts and Telecommunications, Beijing, China \and
Centre for Medical Imaging, University College London, London, UK \and
National Institute of Health Data Science, Peking University, Beijing, China\\
\email{xuyinsong@bupt.edu.cn, yipeng.hu@ucl.ac.uk}}
\authorrunning{Y. Xu et al.}
\maketitle              % typeset the header of the contribution
\begin{abstract}
The Gleason groups serve as the primary histological grading system for prostate cancer, providing crucial insights into the cancer’s potential for growth and metastasis. In clinical practice, pathologists determine the Gleason groups based on specimens obtained from ultrasound-guided biopsies. In this study, we investigate the feasibility of directly estimating the Gleason groups from MRI scans to reduce otherwise required biopsies. We identify two characteristics of this task, ordinality and the resulting dependent yet unknown variances between Gleason groups. In addition to the inter-/intra-observer variability in a multi-step Gleason scoring process based on the interpretation of Gleason patterns, our MR-based prediction is also subject to specimen sampling variance and, to a lesser degree, varying MR imaging protocols. To address this challenge, we propose a novel Poisson ordinal network (PON). PONs model the prediction using a Poisson distribution and leverages Poisson encoding and Poisson focal loss to capture a learnable dependency between ordinal classes (here, Gleason groups), rather than relying solely on the numerical ground-truth (e.g. Gleason Groups 1-5 or Gleason Scores 6-10). To improve this modelling efficacy, PONs also employ contrastive learning with a memory bank to regularise intra-class variance, decoupling the memory requirement of contrast learning from the batch size. Experimental results based on the images labelled by saturation biopsies from 265 prior-biopsy-blind patients, across two tasks demonstrate the superiority and effectiveness of our proposed method. The source code is available at \href{https://github.com/Yinsongxu/PON.git}{https://github.com/Yinsongxu/PON.git}.
\end{abstract}

\section{Introduction}
\label{sec:intro}

\begin{figure}
    \centering
    \includegraphics[width=1\textwidth]{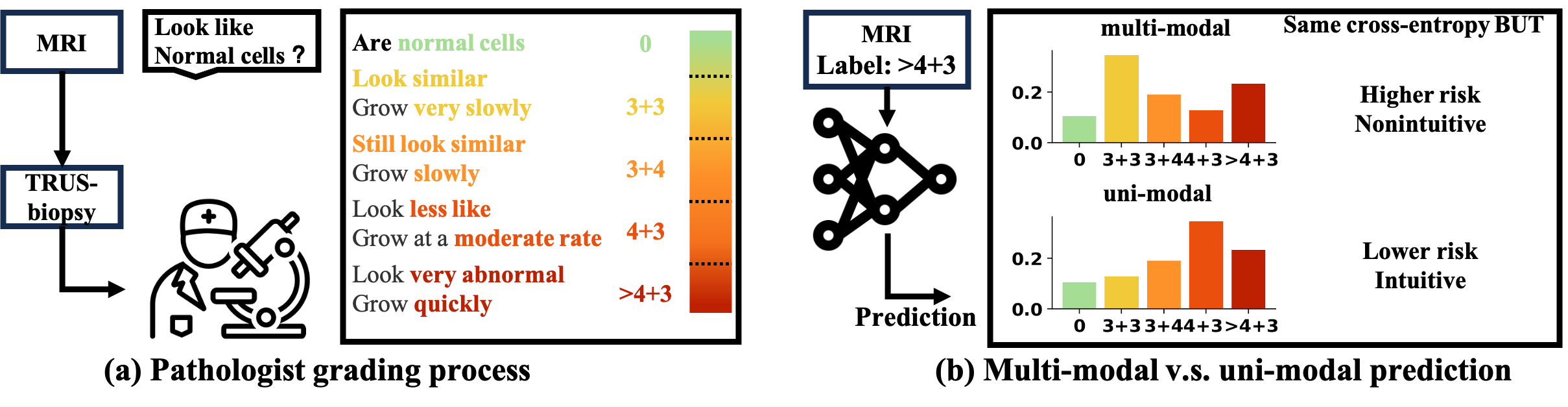}
    \caption{(a) Pathologists assign Gleason groups based on how the cancer cells look like healthy tissue under the microscope. Although cell growth is a continuous process, pathologists discretize it (dotted lines) into multiple groups. (b) Both predictions are incorrect and yield the same cross-entropy loss. However, a multi-modal distribution tends to produce misclassifications that deviate further from the label, resulting in different treatments. Consequently, it poses a higher clinical risk. Consider an image predicted as \texttt{3+3}. The multi-modal distribution appears counter-intuitive, as pathologists will not assign second highest confidence to \texttt{>4+3} due to the obvious difference.}
    \label{fig:teaser}
\end{figure}

Prostate cancer (PCa) is the most frequently diagnosed malignancy in 105 countries \cite{bray2018global}, potentially contributed by both its high prevalence and over-diagnosis using previous blind transrectal ultrasound (TRUS)-guided biopsy.
As illustrated in Fig. \ref{fig:teaser}(a), the current diagnostic process has improved the TRUS biopsy, by adopting magnetic resonance (MR) imaging as a triage test before biopsy \cite{ahmed2017diagnostic}. However, this MR-targeted biopsy is subject to a significant (albeit much lower than blind biopsies) false-positive rate which leads to considerable unnecessary biopsies in men without cancer. In addition to the inherent specificity of MR imaging, the current false-positive rate may also be explained by the complexity and subjectivity in the biopsy and subsequent histology examination procedures. 

Following the biopsy, the cancer-severity-indicating Gleason groups (on a scale of 1-5) are assigned to each biopsy-obtained tissue sample, during pathological examination~\cite{stark2009gleason}. In this study, we discuss the ordinal nature of this classification system and highlight that the dependency between Gleason groups is highly variable and yet quantified, which may be attributed to a lack of studies for predicting Gleason groups directly in literature.
%Consequently, our objective is to advance the field by investigating the potential of directly estimating Gleason grade from MRI using deep neural networks.
Deep learning approaches have been proposed for detecting and segmenting clinically significant lesions from MR images, based on radiologist labels \cite{YAN2024103030}, targeted-biopsy labels~\cite{saha2023artificial} and on labels obtained from prostatectomy patients - which is subject to a different set of challenges such as shifting patient cohort and registration~\cite{cao2019joint}, among other relevant tasks, e.g. needle placement \cite{gayo2022strategising} and registration \cite{shen2022collaborative}. 
%Despite these advancements, Gleason grade estimation remains an underdeveloped area. 
%We articulate and elucidate the key characteristics of this challenging task, and model it as an ordinal classification task with large intra-class variance. 

\textbf{Ordinal classification and class dependency.} 
%Cancers characterized by higher Gleason groups exhibit increased aggressiveness and a worse prognosis. Unlike discrete classification, where classes are orthogonal, Gleason group estimation operates on an ordinal scale. 
Gleason Groups 1-5 are defined with an inherent order, indicating increasingly aggressive cancer which predicts a poorer prognosis. Fig.\ref{fig:teaser}(b) illustrates an example of problems using one-hot encoding with cross-entropy loss, which assumes independence between classes. For instance, given a ground-truth label \texttt{>4+3}, the provided two examples, which would be classified as \texttt{3+3} and \texttt{4+3}, lead to the same loss values (indicating the same risks) if the class-independent encoding is used. However, the former mistake (classified as \texttt{3+3}) has a more serious consequence, which should have been indicated by a higher risk. 

Ordinal encoding is considered an efficient label representation for this type of task \cite{potdar2017comparative}, alleviating the above-discussed issue. Furthermore, we also investigate modelling the change between classes with an uni-modal distribution (e.g. Poison described in Sec.~\ref{op}), for incorporating a prior on the dependency between these ordered classes. 
%With further reasoning discussed in Sec.XXX, we demonstrate an illustrative examples. Given the input MRI, we compare the two classification predictions from neural networks in . Both predictions exhibit the same cross-entropy loss under one-hot target encoding, indicating similar performance for a neural network. 
%It can not reflect they have similar clinical performance. Cancers with adjacent groups may receive similar treatments. 
This uni-modal assumption reflects the fact that, in practice, it is more likely to have class changes (the random variable in the example Poison distribution models) near the ``middle'' classes (in this case, \texttt{3+3}$\rightarrow$\texttt{3+4} and \texttt{3+4}$\rightarrow$\texttt{4+3}) than at two ends (e.g. \texttt{4+3}$\rightarrow$\texttt{>4+3}). This is also echoed with clinical practice in which more middle-class cancers are considered equivocal, therefore associating a higher possibility of changing their labels.
%For instance, both \texttt{4+3} and \texttt{>4+3} cancers are clinically significant. In this case, the unimodal distribution, which misclassifies \texttt{>4+3} as \texttt{4+3}, shows lower risk than the multimodal distribution, which predicts \texttt{3+3}.

%Furthermore, it diverges from regression due to the fact that there is no explicit clinical metric (\eg weight, volume) for measuring distances between groups. Thus, we consider Gleason group estimation as an ordinal classification task.

%For example, pathologists may struggle between two adjacent groups and implausibly assign confidence like multi-modal distribution, which assigns the highest confidence to \texttt{3+3} and the second highest confidence to \texttt{>4+3}. In summary, a unimodal distribution on the change between classes aligns better with the ordinal characteristic of Gleason groups.

\textbf{Unknown inter- and intra-class distributions.} 
Furthermore, the Gleason groups have undergone two discretisation steps, assigning one of the five Gleason Patterns to represent a continuously changing pattern at the cellular scale, before being grouped into the five Gleason groups by ranking majority and secondary Gleason patterns \cite{stark2009gleason}. 
For example, Gleason groups are assigned based on the appearance characterising Gleason patterns on obtained tissue specimens~\cite{stark2009gleason}, as depicted in Fig.\ref{fig:teaser}(a). Patients may exhibit multiple groups of cells with varying proportions and locations. MRIs within the same group can display diverse appearances, leading to high intra-class variance as well as the dependent inter-class variance. This results in a complex unknown intra-/inter-class covariance, subject to subjectivity and uncertainty in either step for our application, and also to the biopsy sampling and imaging. This may challenge the design of the ordinal encoding with a predefined and/or fixed class distribution.

% For instance, both \texttt{4+3} and \texttt{>4+3} cancers are clinically significant. The uni-modal distribution which misclassify \texttt{>4+3} as \texttt{4+3} shows lower risk than multi-modal distribution which predict \texttt{3+3}.

Motivated by the aforementioned observations, we propose the \textbf{P}oisson \textbf{O}rdinal \textbf{N}etwork (\textbf{PON}) for Gleason group estimation. First, we model the prediction distribution using a Poisson distribution. Instead of normalization (such as softmax) on the classifier’s output as prediction distribution, we employ the output as the learnable Poisson's distribution parameter. We then introduce Poisson encoding, which assigns values based on the learnable Poisson distribution. For supervision, we minimise the Kullback-Leibler (KL) divergence between the predicted distribution and the encoded labels. It offers two key advantages. First, it encourages the network to learn the inherent order of classes. Second, it prompts the model to adapt the label distribution rather than the numerical ground-truth. Finally, we further propose a memory-bank-based contrastive learning to enhance representation by regulating the inter-/intra-class covariance.

Our contributions are threefold. 1) We introduce a novel Poisson encoding approach to model ordinal Gleason groups. 2) We propose the PONs, a novel ordinal classification network and its training strategy for the challenging Gleason group estimation.  3) We evaluate our method across two tasks, on a densely-labelled clinical bi-parametric MR data set, demonstrating its superiority over existing state-of-the-art approaches.

\section{Methods}

\begin{figure}
    \centering
\includegraphics[width=1\textwidth]{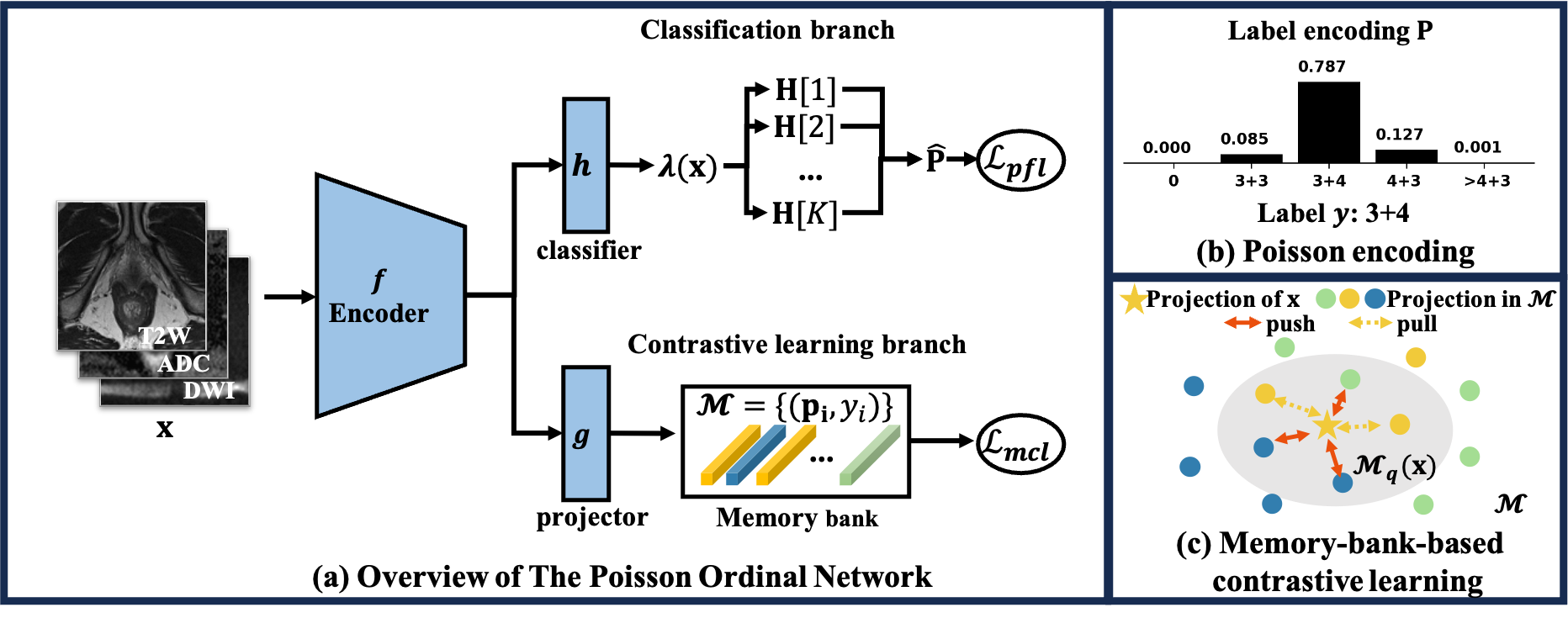}
    \caption{(a) Our method represents the prediction probability using a Poisson distribution with the parameter $\lambda(\textbf{x})$ output from the classifier, and stores the output from projector in a memory bank. (b) For ordinal classification, we introduce the inherent class ordering through Poisson encoding. (c) To regulate the inter-/intra-class covariance, we conduct contrastive learning between the sample and the memory bank, which pulls together features with the same label and pushes away different labeled features}
    \label{fig:overall}
\end{figure}

\subsection{Overview}
The Poisson Ordinal Network (PON), depicted in Fig.\ref{fig:overall}(a), consists of an encoder $f(\cdot)$ followed by two parallel branches: a classification branch and a contrastive learning branch. In the classification brunch, we model prediction distribution as a Poisson distribution with parameters derived from the output of classifier $h(\cdot)$. Then, we propose Poisson encoding to introduce the inherent class ordering into supervision, as shown in Fig.\ref{fig:overall}(b), and the Poisson focal loss, $\mathcal{L}_{pfl}$ to adjust the loss contributions for different samples. In the contrastive learning branch, the projection output from the projector $g(\cdot)$ is stored in the memory bank $\mathcal{M}$. We conduct contrastive learning between the projection of input and those stored in the memory bank to improve the representation ability of the encoder as shown in Fig.\ref{fig:overall}(c). The details are described as follows.

% Given an image $\textbf{x}$, the encoder first extracts its feature. The classifier outputs the parameter $\lambda(\textbf{x}) =\sigma \circ h \circ f (\textbf{x})$ of Poisson distribution for ordinal prediction ($\sigma(\cdot)$ is softplus function to enforced $\lambda(\textbf{x})>0$). Meanwhile, $\lambda(\textbf{x})$ is also be utilized in sample-wise regularization to enhance noise robustness. The projector outputs a projection $\textbf{p} = g \circ f (\textbf{x})$ which is stored in in memory bank $\mathcal{M}$ to reduce intra-variance by clustering. 
%However, in ordinal classification scenarios, it only "cares" about the ground truth class while ignores the inherent ordering between the classes. 

\subsection{Poisson-based Prediction}
\label{op}
We enforce uni-modal constraints by modelling the predicted probability using a Poisson distribution \cite{beckham2017unimodal}. The Poisson distribution describes the probability of a specific number of events occurring within a fixed time interval. For a discrete random variable $v$, the probability mass function with parameter $\lambda>0$ is: 

\begin{equation}
\text{Pr}(v=k) = \frac{\lambda^k \exp(-\lambda)}{k!},
\end{equation}
where $k$ is the number of occurrences ($k=0, 1, 2, \dots$ ).

For Gleason group estimation, we consider an advancing by 1 group as the ``event'', \eg transitioning from \texttt{3+4} to \texttt{4+3}. Consequently, the number of events corresponds to the class index (0-4). To model this, we replace the $\lambda$ with the scalar output of classifier $\lambda(\textbf{x}) = h \circ f (\textbf{x})$, where $\circ$ denotes function composition ($\lambda(\textbf{x})\in \mathbbm{R}^+$ is enforced by softplus function). Considering the $K$-class classification, we obtain the predicted distribution $\hat{\textbf{P}} \in \mathbbm{R}^K$ after normalization:

\begin{equation}
\label{lbd}
\hat{\textbf{P}}[k] =  \frac{\lambda(\textbf{x})^k \exp(-\lambda(\textbf{x}))/k!}{\sum_{k=1}^K \lambda(\textbf{x})^k \exp(-\lambda(\textbf{x}))/k!}.
\end{equation}

For ease of implementation, we introduce the auxiliary variable $\textbf{H} \in \mathbbm{R}^K$:
\begin{equation}
\textbf{H}[k] = \log[\lambda(\textbf{x})^k \exp(-\lambda(\textbf{x}))/k!] = k\log(\lambda(\textbf{x})) - \lambda(\textbf{x}) -\log(k!).
\end{equation}
In this way, $\hat{\textbf{P}}$ can be derived using the softmax function applied to $\textbf{H}$.

% \begin{equation}
% p(y=k|\textbf{x}) = \frac{\exp(h(\textbf{x})_k)}{\sum_{k=1}^K \exp(h(\textbf{x})_k)}.
% \end{equation}

\subsection{Poisson Encoding and Poisson Focal Loss}
\label{pfl}

The commonly-used combination of one-hot encoding and cross-entropy does not account for the inherent order of classes or explicitly encourage models to follow uni-modal distributions, as discussed in Sec.~\ref{sec:intro}. To address this limitation, we propose a novel approach called Poisson encoding. Formally, we transform the label $y$ to a Poisson distribution $\textbf{P}\in \mathbbm{R}^K$:
\begin{equation}
\textbf{P}[k] = \frac{(y^k \exp(-y)/k!)^t}{\sum_{k=1}^K (y^k \exp(-y)/k!)^t},
\end{equation}
where $t$ is the temperature hyperparameter to control the smoothness of distribution. As a supervision, we introduce Poisson focal loss, $\mathcal{L}_{pfl}$, which adapts focal loss \cite{lin2017focal} to Poisson encoding for ordinal classification:
\begin{equation}
\mathcal{L}_{pfl} = -(\textbf{P}[y] - \hat{\textbf{P}}[y])^{\gamma} \text{KL}(\textbf{P}||\hat{\textbf{P}}),
\end{equation}
where $\gamma \geq 0$ serves as the focusing parameter, and $\text{KL}(\cdot||\cdot)$ is Kullback-Leibler divergence. When a sample is accurately classified, $(\textbf{P}[y] - \hat{\textbf{P}}[y])^{\gamma}$ approaches $0$, contributing minimally to the overall loss. Consequently, the training emphasizes challenging samples. To summarise, the Poisson focal loss offers two key advantages over the with vanilla focal loss \cite{lin2017focal} and cross entropy: (1) it introduces the inherent order of classes through label encoding. (2) it encourages the modal to consider not only the expected numerical ground-truth class but also the distribution of the predictions.

\subsection{Memory-bank-based Contrastive Learning}
\label{cl}

Taking into consideration of the diverse appearances of MRIs within the same group, discussed in Sec.~\ref{sec:intro}, we propose contrastive learning to improve the representation learning. However, existing methods often heavily depend on batch size to enhance the range of contrast \cite{wang2021contrastive,zhang2023ecl}. When dealing with 3D MRI data, the memory constraints associated with small batch sizes may introduce bias towards samples in the mini-batch. To decouple the contrastive learning from the batch-size, we introduce memory-bank-based contrastive learning. Formally, we propose the memory bank denoted as $\mathcal{M}$, which stores the projections of all samples $\textbf{x}_i$, $\textbf{p}_i = g\circ f(\textbf{x}_i)$, along with their labels $y_i$, \ie $\mathcal{M}=\{(\textbf{p}_i, y_i)\}^N_{i=1}$, where $N$ is dataset size. We update $\mathcal{M}$ at each forward iteration. Given an image $(\textbf{x}, y$), we select the $q$ nearest elements to $ g\circ f(\textbf{x})$ in $\mathcal{M}$ denoted as $\mathcal{M}_q(\textbf{x})\subseteq  \mathcal{M}$. We propose $\mathcal{L}_{mcl}$ to facilitate contrastive learning:
\begin{equation}
\mathcal{L}_{mcl} = - \frac{\sum_{(\textbf{p}_i,y_i)\in \mathcal{M}_q(\textbf{x})}\exp sim(\textbf{p}_i,  g\circ f(\textbf{x}))\mathbbm{1}\{y_i=y\}}{\sum_{(\textbf{p}_i,y_i)\in \mathcal{M}_q(\textbf{x})}\exp sim(\textbf{p}_i,  g\circ f(\textbf{x}))},
\end{equation}
where $sim(\cdot,\cdot)$ is cosine similarity. With this configuration, features with the same label are pulled together and different labeled features are pushed away.

PONs could be trained in an end-to-end manner with the overall objective:$\mathcal{L} = \mathcal{L}_{pfl} + \mathcal{L}_{mcl}.$
% Finally, we briefly review existing theoretical insights on the optimal choice of regularization strength.

% Generally, the optimal regularization strength for a given model family increases with the label noise level and decreases in the sample size. 
\section{Experiments}
\subsection{Experiment Setup}

\textbf{Dataset.}
Our method is evaluated on the PROMIS dataset \cite{ahmed2017diagnostic} consisting of 265 publicly available MRIs from 262 patients. The Gleason Scores, obtained through template prostate mapping biopsy\cite{valerio2016transperineal}, were determined by expert uropathologists based on core biopsies taken at 5 mm intervals. Among all patients, 67 patients have no cancer, 49, 89, 31, and 26 patients were reported with Gleason group \texttt{3+3}, \texttt{3+4}, \texttt{4+3}, and \texttt{>4+3}, respectively. We utilize 3 image volumes from bi-parametric sequences, including T2-weighted (T2W), diffusion-weighted with high b-value (DWI), and apparent diffusion coefficient (ADC). All sequences are resampled to voxels of $0.5\times 0.5 \times 2$ mm$^3$ with trilinear interpolation. All sequences are aligned with T2W images as the reference, and cropped into a $128 \times 128 \times32$ region of interest (ROI) centred at the prostate’s centroid.

\noindent \textbf{Evaluation Metrics.}
We evaluate two tasks: Gleason group prediction (five classes), and detection of clinically significant cancer (primary and secondary definitions). For all tasks, we utilize five-fold cross-validation. Each experiment is repeated five times with different random seeds, and we report the mean and standard deviation. We adopt the following metrics for evaluation: accuracy (Acc), macro area under curve (AUC), quadratic weighted kappa (QWK), F1-score, sensitivity at specificity(Sen@Spec), and specificity at sensitivity(Spec@Sen).

\noindent \textbf{Implementation Details.}
We employ the 3D ResNet18 architecture as the encoder. The classifier consists of a fully connected layer, and the projector is a two-layer multilayer perceptron. For optimization, we adopt the Adam optimizer with an initial learning rate of 1e-4, training for 60 epochs. All experiments are conducted using PyTorch on RTX6000 GPUs. The hyperparameters $t$, $\gamma$, and $q$ are set to 0.1, 2 and 20, respectively. To keep a balanced number of samples for each class, we utilise a weighted sampler during training.

% We evaluate two tasks with clinical significance: risk group(no cancer, Gleason Score $\leq6$, $=7$, $>7$) and significant detection (no cancer and Gleason score $< 3+4$ vs Gleason $\leq 3+4$). We employ the five-fold cross-validation for all tasks. 

% For risk group, as it is ordinary classification, we propose accuracy and  the quadratic weighted kappa (QWK) \cite{cohen1968weighted} as metric.

% Ordinal encoding with FocalLoss, FocalNet \cite{cao2019joint

\begin{table*}[t]
\centering
\small
\caption{Results of Gleason group prediction. (mean$\pm$standard)}
\label{tab1}
\begin{tabular*}{\hsize}{@{}@{\extracolsep{\fill}}lcccc@{}}
\hline
Method& Acc$\uparrow$& AUC$\uparrow$& QWK$\uparrow$& F1$\uparrow$\\\hline
CE& 28.08$\pm$2.33& 53.91$\pm$0.61& 0.03$\pm$0.01& 20.62$\pm$1.49\\
FocalLoss \cite{lin2017focal}& 30.70$\pm$1.81& 55.02$\pm$1.54& 0.06$\pm$0.03& 22.82$\pm$1.89\\
EMD \cite{hou2016squared}& \textbf{31.47}$\pm$2.10& 57.84$\pm$2.58& \textbf{0.08}$\pm$0.04& 25.03$\pm$2.18\\
ExpLoss \cite{qin2022ultra}& 31.15$\pm$0.62& 56.78$\pm$1.89& \textbf{0.08}$\pm$0.01& 24.67$\pm$1.45\\
OrdinalEncoding \cite{potdar2017comparative}& 29.49$\pm$1.60& - & 0.06$\pm$0.02& 22.91$\pm$2.38\\
Softlabel\cite{potdar2017comparative}& 25.13$\pm$3.66& 56.19$\pm$3.40& 0.05$\pm$0.03& 22.39$\pm$3.36\\
Ours& 29.35$\pm$1.82& \textbf{60.06}$\pm$1.22& 0.07$\pm$0.02& \textbf{25.58}$\pm$2.25\\ \hline
\end{tabular*}
\end{table*}

\begin{table}[t]

\centering
\small
\caption{Results of detection of clinically significant cancer (mean$\pm$standard).}
\label{tab2}
\begin{tabular*}{\hsize}{@{}@{\extracolsep{\fill}}lcccc@{}}
\hline
Method& Spec@Sen80\%& Sen@Spec80\%& Spec@Sen90\%& Sen@Spec90\%\\\hline
\multicolumn{5}{c}{Primary definition: Gleason score $\geq$ 4+3}  \\
\hline
CE& 32.46$\pm$5.31& 31.76$\pm$3.55& 15.74$\pm$5.35& 16.77$\pm$5.47\\
FocalLoss\cite{lin2017focal}& 34.02$\pm$5.20& 38.53$\pm$7.64& 19.34$\pm$5.64& 21.47$\pm$6.61\\
EMD \cite{hou2016squared}& 38.77$\pm$7.02& 39.12$\pm$3.56& 23.44$\pm$7.27& 24.12$\pm$2.56\\
ExpLoss \cite{qin2022ultra}& 35.08$\pm$7.62& 36.47$\pm$5.77& 21.97$\pm$8.23& 21.77$\pm$3.00\\
Softlabel\cite{potdar2017comparative}& 40.82$\pm$3.68& 31.17$\pm$1.71& 24.26$\pm$8.32& 15.59@9.33\\
Ours& \textbf{43.60}$\pm$6.06& \textbf{44.12}$\pm$6.77& \textbf{27.62}$\pm$8.80&\textbf{26.18}$\pm$5.91\\ \hline
\multicolumn{5}{c}{Secondary definition: Gleason score $\geq$ 3+4}  \\
\hline
CE& 41.51$\pm$3.93& 36.00$\pm$3.77& 26.64$\pm$5.15& 22.45$\pm$2.57\\
FocalLoss\cite{lin2017focal}& 44.70$\pm$4.98& 36.33$\pm$6.64& 28.33$\pm$4.20& 23.11$\pm$6.97\\
EMD \cite{hou2016squared}& \textbf{45.91}$\pm$1.03& 36.44$\pm$7.84& \textbf{30.91}$\pm$3.08& 20.44$\pm$4.99\\
ExpLoss \cite{qin2022ultra}& 44.39$\pm$3.44& 38.78$\pm$5.29& 26.06$\pm$4.20& 22.67$\pm$2.66\\
Softlabel\cite{potdar2017comparative}& 40.00$\pm$2.22& 37.00$\pm$5.64& 25.46$\pm$3.31& 17.24$\pm$8.66\\      
Ours& 43.79$\pm$8.76& \textbf{39.00}$\pm$7.32& 26.51$\pm$7.22& \textbf{25.00}$\pm$5.60\\
\hline
\end{tabular*}
\end{table}

\subsection{Results}
\noindent \textbf{Comparison Results.} To evaluate the performance of PON, we compare our method with ordinal classification methods, including CrossEntropy (CE), re-weighting methods (FocalLoss \cite{lin2017focal}), ordinal classification loss (Earth mover's distance, EMD\cite{hou2016squared},
and ExpLoss \cite{qin2022ultra}), and encoding methods (OrdinalEncoding \cite{potdar2017comparative}, Softlabel\cite{potdar2017comparative}). To ensure fairness, we reproduce all methods on our datasets with the same experimental settings. The results of Gleason group prediction and detection of clinically significant cancer are shown in Tab.\ref{tab1} and \ref{tab2}, respectively. It can be seen that PON shows a significant advantage with the highest performance in most metrics. Noticeably, PON outperforms other methods by great gains in the detection of primary definition significant cancer \eg, outperforms EMD 4.83\% in Spec@Sen80\%, and 5.00\% in Sen@Spec 80\% (p-value=1.58e-3 with the Mann–Whitney U test).

\begin{table*}[t]
\centering
\small
\caption{Ablation study. PP: Poisson-based Prediction, PE:Poisson encoding.}
\label{tab:abl}
\begin{tabular*}{\hsize}{@{}@{\extracolsep{\fill}}cccccccc@{}}
\hline
PP& PE& $\mathcal{L}_{pfl}$& $\mathcal{L}_{mcl}$& Acc$\uparrow$& AUC$\uparrow$& QWK$\uparrow$& F1$\uparrow$\\\hline
&&&&28.08$\pm$2.33& 53.91$\pm$0.61& 0.03$\pm$0.01& 20.62$\pm$1.49\\
\checkmark& & & & 27.37$\pm$1.86& 59.04$\pm$0.95 & 0.04$\pm$0.02& 21.17$\pm$2.55\\
\checkmark& \checkmark& & & 29.62$\pm$2.13& 60.43$\pm$1.49& \textbf{0.07}$\pm$0.02&22.92$\pm$1.90\\
\checkmark& \checkmark& \checkmark& &28.85$\pm$1.93& 60.41$\pm$2.49& \textbf{0.07}$\pm$0.02& 24.14$\pm$1.09\\
& & & \checkmark& \textbf{30.51}$\pm$2.89& 55.36$\pm$0.62& 0.06$\pm$0.03& 23.18$\pm$2.15\\
\checkmark & \checkmark &\checkmark  &\checkmark & 29.35$\pm$1.82& \textbf{60.06}$\pm$1.22& \textbf{0.07}$\pm$0.02& \textbf{25.58}$\pm$2.25\\ \hline
\end{tabular*}
\end{table*}

% mem only
% 35.58, 56.48, 0.10442134125442715, 75.00, 62.95, 82.37, 55.78
% 31.41, 52.78, 0.06725154368416641, 72.44, 62.47, 83.33, 55.47
% 29.81, 57.79, 0.010613805186719172, 75.64, 56.27, 88.46, 70.13
% 31.73, 55.91, 0.06485611763878141, 71.47, 60.11, 81.09, 62.28
% 34.94, 56.65, 0.09233436993938005, 74.04, 64.71, 87.50, 63.50

%poisson only
%28.53, 60.85, 0.06674535894409261, 70.83, 68.78, 83.97, 71.56
%26.92, 58.95, 0.04456442904343627, 71.79, 70.77, 83.33, 64.40
%31.09, 60.80, 0.08172484599589336, 74.36, 70.63, 80.45, 71.10
%25.96, 59.49, 0.034366333051971565, 70.19, 65.21, 80.13, 61.36
%26.28, 61.80, 0.042574481994903324, 70.19, 69.22, 82.69, 71.86

%poisson + PFL
%31.73, 63.34, 0.08904485140914598, 71.15, 71.21, 87.82, 69.63
%31.09, 63.89, 0.103819588248654, 73.40, 71.49, 88.78, 72.18
%29.49, 62.08, 0.06472271426624887, 74.36, 71.75, 88.14, 73.77
%34.62, 62.53, 0.13834323852329178, 73.72, 69.79, 84.29, 70.98
%26.92, 62.44, 0.04327944696990049, 72.44, 66.01, 89.42, 77.19

%poisson mem ce
%34.94, 64.88, 0.1322528052172247, 74.68, 76.18, 85.26, 69.00
%32.69, 61.04, 0.11776586863436822, 73.08, 67.96, 86.86, 68.44
%30.77, 62.42, 0.08677977126131498, 72.76, 67.19, 86.86, 75.57
%29.17, 61.28, 0.07926503578677502, 70.83, 69.02, 83.97, 71.45
%26.92, 59.70, 0.05055789867065286, 70.51, 64.16, 77.88, 68.30

% \input{table/rad}
\begin{figure}
    \centering
\includegraphics[width=0.8\textwidth]{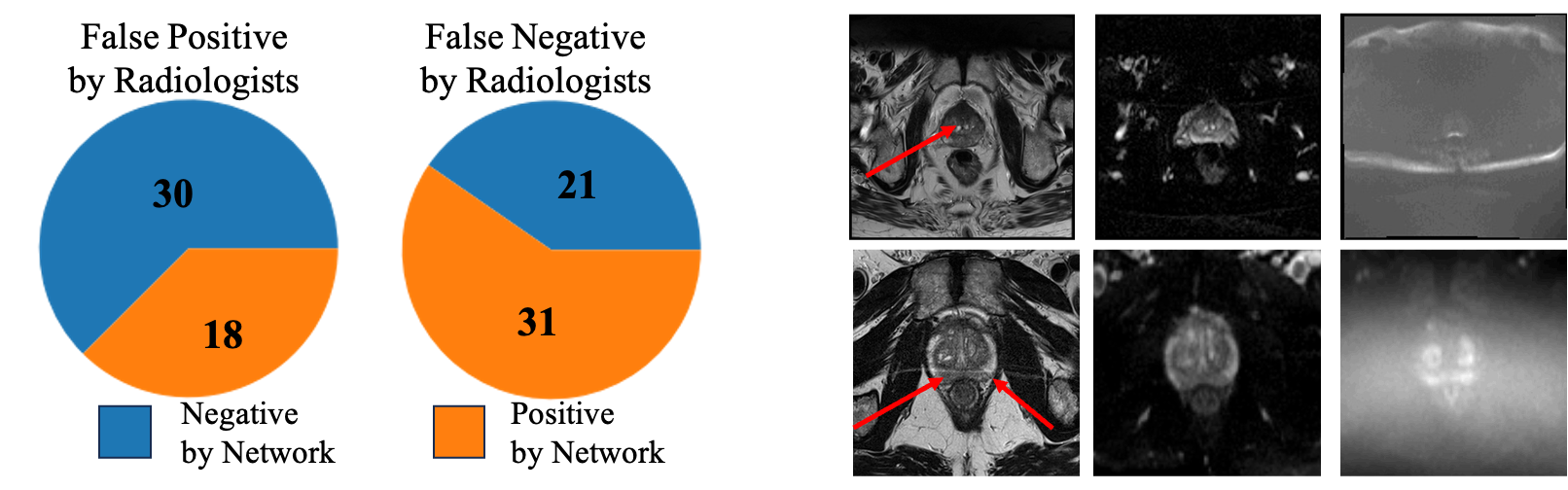}
    \caption{\textbf{Left}: our method helps distinguish 30 negatives in 48 false positive samples under the same sensitivity, and 31 positive samples in 52 false negative samples under the same specificity by radiologists. \textbf{Right}: Two examples of the network predicting correctly and radiologists predicting wrong. Arrows indicate lesions.}
    \label{fig:ana}
\end{figure}

\noindent \textbf{Ablation Study.} To verify the impact of each component in the proposed PON, we conducted an ablation study, as detailed in Tab.\ref{tab:abl}. We propose cross-entropy as the baseline. Notably, the inclusion of Poisson-based predictions and Poisson encoding significantly enhances the AUC and QWK, \eg a considerable 5.13\% improvement in AUC. It demonstrates the effectiveness of uni-modal constraints on prediction distributions, enhancing the model’s ability to learn from ordinal data. 
Furthermore, when combined with $\mathcal{L}_{pfl}$, F1-score further improves. Subsequently, we introduced $\mathcal{L}_{mcl}$ to the baseline. The improvement suggests that this strategy can significantly enhance the network’s data representation ability by regulating  inter-/intra-class covariance. In summary, PON outperforms other models across most evaluation metrics, highlighting its suitability for ordinal classification tasks characterized by large intra-class variance.

\noindent \textbf{Comparison with Radiologists on clinically significant cancer detection.} 
MRI scans were reported at each centre by experienced urologic radiologists prior (and blind) to obtaining densely sampled histopathology labels~\cite{ahmed2017diagnostic}. This provides a unique opportunity to compare the machine-predicted with radiologists' ability in grading prostate cancer, with histopathology as ground-truth. Radiologists achieved sensitivity 75.82\%/63.63\%, and specificity 75.00\%/71.11\%, for primary and secondary definitions respectively. We compare our method with radiologists by adjusting the threshold to control sensitivity and specificity. Although the network did not outperform expert radiologists, it can assist radiologists' decision-making. As shown in Fig.\ref{fig:ana}-left, the network can distinguish 30 negative samples in a total of radiological 48 false-positives with the same sensitivity, and 31 positive samples in 52 radiological false-negatives with the same specificity, quantitatively confirming its added clinical value. Two example cases in which the machine predictions can be used to co-pilot radiologists, either as a first- or second reader, are provided in Fig.\ref{fig:ana}-right.

\section{Conclusion}
In this work, we present a Poisson ordinal network (PON) for Gleason group estimation. We identify two characteristics of this task, ordinality and a dependent and unknown variance between Gleason groups. To address this challenge, PONs models the prediction using a Poisson distribution and leverages Poisson encoding and Poisson focal loss to capture a learnable dependency between ordinal classes.  Furthermore, to improve this modelling efficacy, PONs also employ contrastive learning with a memory bank. Experimental results based on the images labelled by saturation biopsies across two tasks demonstrate the superiority and effectiveness of our proposed method. 

\begin{credits}
\subsubsection{\ackname}
This work was supported by the International Alliance for Cancer Early Detection, a partnership between Cancer Research UK [C28070/A30912; C73666/A31378], Canary Center at Stanford University, the University of Cambridge, OHSU Knight Cancer Institute, University College London and the University of Manchester. This work was also supported by the China Scholarship Council.
\subsubsection{\discintname}
The authors have no competing interests to declare that are relevant to the content of this article.
\end{credits}

%
%

\bibliographystyle{splncs04}
\bibliography{Paper-0193}

\end{document}